\documentclass[quaderni]{sifrev}        
\usepackage{amsmath}
	\renewcommand\and{\mbox{\rm and\ }\ignorespaces}

\usepackage{graphicx}
\usepackage[applemac]{inputenc}
\usepackage{fontenc}

\begin{document}
\title{{Galileo Galilei and a forgotten poem on the 1604 supernova}}

\shorttitle{Galileo Galilei and a forgotten poem on  SN1604}

\author{Alessandro De Angelis\thanks{E-mail: alessandro.deangelis@unipd.it}}           

\institute{Dipartimento di Fisica e Astronomia ``Galileo Galilei'' dell'Universit\`a di Padova\\
Istituto Nazionale di Fisica Nucleare \& INAF, Padova, Italy\\
Universidade de Lisboa \& Instituto Superior T\'ecnico/LIP,  Lisboa, Portugal}

\abstract{
In October 1604, when SN1604, the last naked-eye visible supernova  in our Galaxy, exploded, Galileo Galilei was  professor of mathematics   at the University of Padua, teaching the mechanics of planets. He was therefore the figure of reference to whom all the doubts and questions that such an apparition brought with it were addressed. The University   asked Galilei to outline the situation by exposing in three public conferences his point of view, in order to answer the many questions that raged among the academic community and the common people. Three conferences that Galilei held  between November and December, in the Aula Magna of the Bo, the central building of the University. A month after Galilei's lectures, a treatise on the supernova appeared in Padua. The unknown Antonio Lorenzini, behind whose name it is easy to see the inspiration of Cesare Cremonini, an Aristotelian professor of natural philosophy in Padua, published a booklet entitled {\em Discourse about the new star} which debunked the conclusions of Galilei. One month later was published in Padua by the same editor the {\em Dialogue by Cecco di Ronchitti from Bruzene about the new star,} a booklet in Paduan dialect replying to Lorenzini.   The {\em Dialogue} appears in the Italian national edition of Galilei's works and is attributed to Galilei himself with the help of the monk Girolamo Spinelli; it does not appear instead a poem in octaves, by an unknown author (presumably Galilei), published as an appendix to the first edition of 1605, and immediately replaced in the second edition. We publish it here, with a comment.}

\riassunto{Nell'ottobre 1604, quando esplose la supernova SN1604, l'ultima visibile a occhio nudo nella nostra Galassia, Galileo Galilei era professore di matematica   all'Universit\`a di Padova e insegnava la meccanica dei pianeti. Fu quindi la figura di riferimento a cui vennero rivolti tutti i dubbi e le domande che una tale apparizione portava con sé. L'Universit\`a   chiese a Galilei di fare il punto sulla situazione esponendo in tre conferenze pubbliche le sue convinzioni, per rispondere alle tante domande che imperversavano tra la comunit\`a accademica e la gente comune. Tre conferenze che Galilei tenne   tra novembre e dicembre, nell'Aula Magna del Bo, l'edificio centrale dell'Universit\`a. Un mese dopo le conferenze di Galilei, fu pubblicato a Padova un trattato sulla supernova. Lo scono\-sciu\-to Antonio Lorenzini, dietro il cui nome \`e facile scorgere l'ispirazione di Cesare Cremonini, professore aristotelico di filosofia naturale a Padova, pubblic\`o un {\em Discorso sulla nuova stella,} che contestava le conclusioni di Galilei. Un mese dopo fu pubblicato a Padova presso lo stesso editore il {\em Dialogo de Cecco di Ronchitti da Bruzene in perpuosito de la stella nuova,} un libretto in dialetto padovano che rispondeva a quello di  Lorenzini. Il {\em Dialogo} compare nell'edizione nazionale delle opere di Galilei ed \`e attribuito a Galilei stesso con l'aiuto del monaco Girolamo Spinelli; non vi compare invece un poema in ottave, di autore ignoto (presumibilmente Galilei), pubblicato in appendice alla prima edizione del 1605, e subito sostituito nella seconda edizione. Lo pubblichiamo qui, con un commento.}

\maketitle

\begin{figure}[h]
\begin{center}
\includegraphics[width=0.5\linewidth]{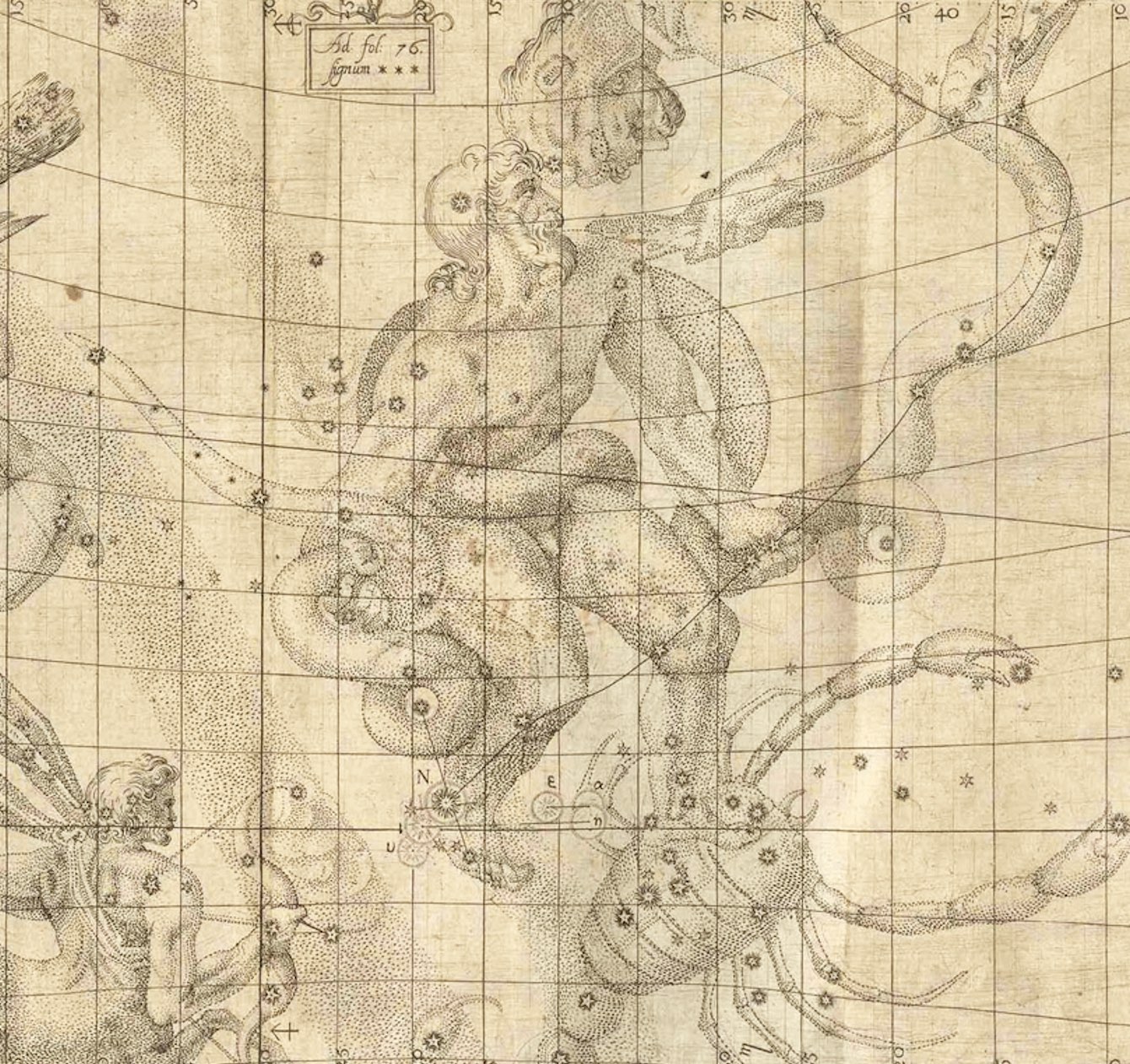}
\caption{\label{kepler}Illustration from Johannes Kepler's book {\em  De stella nova in pede Serpentarii (On the new star in Ophiuchus's foot)}  indicating the location of the 1604 supernova. The supernova is the star marked with a letter ``N'' on the right foot of the Ophiuchus (Serpent Bearer) constellation. Source: Wikimedia commons.}
\end{center}
\end{figure}

On the evening of October 9, 1604, curious and enthusiasts from all over the world were observing a rare conjunction between Mars, Jupiter and Saturn, full of astrological implications. Suddenly, near that place in the sky, which was located in the Ophiucus constellation, near  the foot of the Serpent bearer (Fig. \ref{kepler}), a new light appeared, brighter than all the planets with the exception of Venus. It remained there for a year and a half and then, as it had appeared, it disappeared. That ``new star", as it was called, changed the history of astronomy and cosmology: it was believed at the time that stars were fixed, immutable and ungenerable, but obviously this was not true. Scientists with different conceptions of the Universe competed and collaborated to explain the nature and origin, including Galilei and Kepler, and also Chinese, Korean and Arab astronomers reported observations \cite{universe}. Today we know that that new star was a supernova, the last of the seven supernovae observed with the naked eye in the Milky Way -- the six previous ones of which there is documentation were recorded in the years 185, 393, 1006, 1054, 1181, 1572. So not a new star, but a star that dies and explodes. What remains today is a celestial object still full of mysteries: the cloud projected by the explosion is expanding at very high speed, in some places ten thousand kilometers per second, one thirtieth of the speed of light, and the shock waves of this expansion accelerate cosmic particles up to very high energies. We astrophysicists of today still study it carefully \cite{universe}, always with an eye to the first analysis of our illustrious predecessors of four hundred years ago.

In October 1604, Galileo Galilei was  professor of mathematics and astronomy at the University of Padua, teaching the mechanics of planets. When the {``stella nova''} appeared in the sky, he was therefore the figure of reference to whom all the doubts and questions that such an apparition brought with it  were addressed.

That  bright and pulsating sphere had generated wonder, but also terror and curiosity. The University of Padua asked Galilei to outline the situation by exposing in three public conferences his point of view, in order to answer to the many questions that raged among the academic community and the common people.
Three conferences that Galilei held almost immediately, between November and December, in the Aula Magna of the Bo, the central building of the University -- fragments of these lectures are in the Italian national edition of the Opere di Galileo Galilei (Works of Galileo Galilei), edited by Antonio Favaro  \cite{opere},  vol. II.  Even if today it seems difficult to believe it, Galilei reported that that a thousand people attended each lecture (vol. II of \cite{opere}, \cite{deabio,deasup}). Galilei demonstrated through the method of parallax that the new star was beyond the Moon, and therefore in that Heaven that according to Aristotle was immutable. Parallax is the apparent displacement of an object due to a change in the observer's point of view: looking at the tip of your nose with the left eye closed, and then with the right eye closed, your nose will appear to move; knowing the distance between  eyes one can determine the length of his/her nose. The same thing could be done from different points on the Earth to measure the position of planets and stars, and Galilei did it in collaboration with Neapolitan and Spanish correspondents. The new star was shown to all observers to be at the same location with respect to the stars of Sagittarius and Scorpio, and therefore it had to be farther away than the Moon and the planets: precisely between the fixed stars. Today only a few fragments of Galilei's lecture notes remain (vol. II of \cite{opere}), and they are fascinating -- Galilei was a person of all-round culture, a man of letters as well as a musician, painter and of course a scientist. 

A month after Galilei's lectures, a treatise on the supernova appeared in Padua.  The unknown Antonio Lorenzini, behind whose name it is easy to see the inspiration of Cesare Cremonini, an Aristotelian professor 
of natural philosophy in Padua and friend-rival of his colleague Galilei, published a booklet entitled 
{\em Discourse around the new star} \cite{lorenzini}
which debunked the
conclusions of Galilei. In order to justify Aristotle of Stagira's concept of immutability of the sky, which seemed to be contradicted by the new star, Lorenzini affirmed that looking at the stars to measure their distance was useless: the principles of terrestrial physics did not apply to the sky.

\begin{figure}[h]
\begin{center}
\includegraphics[width=0.5\linewidth]{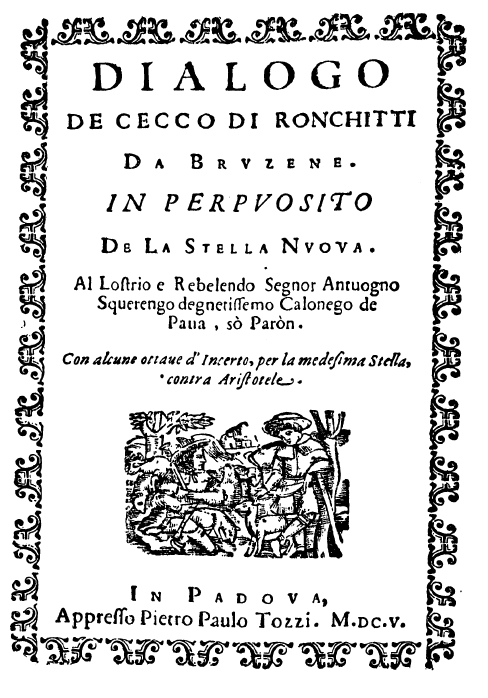}
\caption{\label{cover}The cover page of the first edition of the {\em Dialogo  de Cecco di Ronchitti da Bruzene in perpuosito de la stella nuova.}}
\end{center}
\end{figure}

The response to Lorenzini's speech was  rapid and
original, but also very detailed. On February 28, 1605 was published in Padua, by the same publisher
Tozzi who had published the book of Lorenzini, the {\em Dialogo de Cecco di
Ronchitti da Bruzene in perpuosito de la stella nuova (Dialogue by Cecco di
Ronchitti from Bruzene about the new star)} \cite{bruzene}, a pseudonymous
booklet in Paduan dialect (a local form of Venetian dialect) written
according to Antonio Favaro and to many critics by Galilei  together
with his student and Benedictine monk Girolamo Spinelli (who had a deeper knowledge of the Paduan dialect).  It is not known how
many copies of the book (whose cover page is shown in Fig. \ref{cover}) were published; certainly a second edition was printed the same
year in Verona by the publisher Merlo \cite{bruzene2}, with minimal changes (not
substantial but with some improvements from the point of view of readability) to the text
and a substantial change in the appendix, which contained  {\em Alcune ottave d'incerto, per la medesima stella, contra Aristotele (Some octaves by unknown author, for the same star, against Aristotle)}, in Florentine vernacular, which were substantially modified.

The {\em Dialogo} was an explicit mockery of what had been published by
Lorenzini; it defamed the entire Aristotelian dogma, which explains why the
author(s)   preferred  to hide under the pseudonym of
Cecco di Ronchitti. Cecco di Ronchitti from Brugine (a small village of
peasants near Padua), an unknown self-proclaimed farmer and land
surveyor, gave in the book many indications of being in reality an astronomer
from Padua. The fact that the book was written in Paduan dialect instead of Latin,
which was the noble language used to deal with relevant subjects, was a sign
that the topic around which the booklet was built was not worthy of
consideration,

In addition to the fact that the dialect was perfect for light-hearted teasing, the dialect of Padua was in vogue thanks to the writings of Angelo Beolco, a sort of {\em ante litteram} ``blogger" who had published in the first half of the sixteenth century under the name of ``Ruzante" or ``Ruzzante", which in Paduan means ``scratcher" or ``mumbler", about the events that took place in Padua.

The {\em Dialogue} takes place between two peasants, Matthio (Matteo) and Nale (Natale), who confront each other after Natale read Lorenzini's book and summarizes to Matteo its contents. The booklet analyzes point by point the contradictions and distortions expressed by Lorenzini in his {\em Discourse.}  It is like  Galilei did not want to react personally to what Lorenzini had written, but did so through the dialectal, intelligent and frank voices of two peasants.

It is easy to recognize in the content of the {\em Dialogo} many characteristics of Galilei, so that already at the time of its publication the treatise was attributed to him. In addition to this, no family named Ronchitti is recorded in the birth registers of the parish of Brugine or in the neighboring towns; moreover, Cecco is also a Tuscan nickname.    Galilei had good knowledge of the Paduan vernacular: Niccol\`o Gherardini, in his {\em Vita del signor Galileo Galilei  (Life of Galileo Galilei)} \cite{ghera}, writes: ``he was still very familiar with a book entitled {\em Il Ruzzante,} written in the rustic Paduan language, taking pleasure in those crude tales and ridiculous incidents" (Gherardini confuses the author with the title). Among the {\em Miscellanea  galileiana inedita (Unpublished Galilei's works)} published by Favaro in Venice in 1887 \cite{inedita}, one can read a thought of Galilei that contains words of the rustic Paduan dialect. Finally, in his letter to Paolo Gualdo of June 16, 1612, published in volume XI of the {\em Opere} \cite{opere}, Galilei writes whole sentences using the same dialect, and Galilei's Venetian correspondents often wrote to him in various flavors of the Venetian dialect.
 In a letter to Paolo Gualdo dated August 16, 1614, Girolamo Spinelli was called by Galilei ``one of my students, monk of Santa Giustina, companion of Cecco de' Ronchetti".
 A focus on the form of the text caused Galilei's contribution to be greatly diminished \cite{tomasin} in editions following Favaro's (especially from Paduan critics), in favor of Girolamo Spinelli. This gradual deauthorization has certainly also depended on an excessive attention to form, admittedly so particular, which  has unfortunately come at the expense of attention to specific content, which is equally (if not even more) important \cite{cosci}. 

In the mouth of one of the two interlocutors, Matteo, there are Galilean concepts; we can say that Galilei's ideas about the new star are all in Matteo's words. The Copernican idea of the author of the {\em Dialogue} is in the explanation of the concept of parallax made in the booklet, which also contemplates the seasonal displacement of the Earth linked to its  revolution around the Sun.
Two drawings reproduced from Galilean manuscripts kept at the  Biblioteca Nazionale Centrale di Firenze,  presumably contemporary to the publication of the {\em Dialogue,} and included in a section entitled by Galilei  {\em De stella anni 1604 (About the star of 1604)}, clarify even better the meaning of parallax applied to the Earth's revolution.\footnote{{\em Manoscritti galileiani} 70, c. 16v e 47, c. 10v. Images reproduced by permission of the Italian Ministero della Cultura, Biblioteca
Nazionale Centrale di Firenze.} 

\begin{center}
\includegraphics[width=0.6\linewidth]{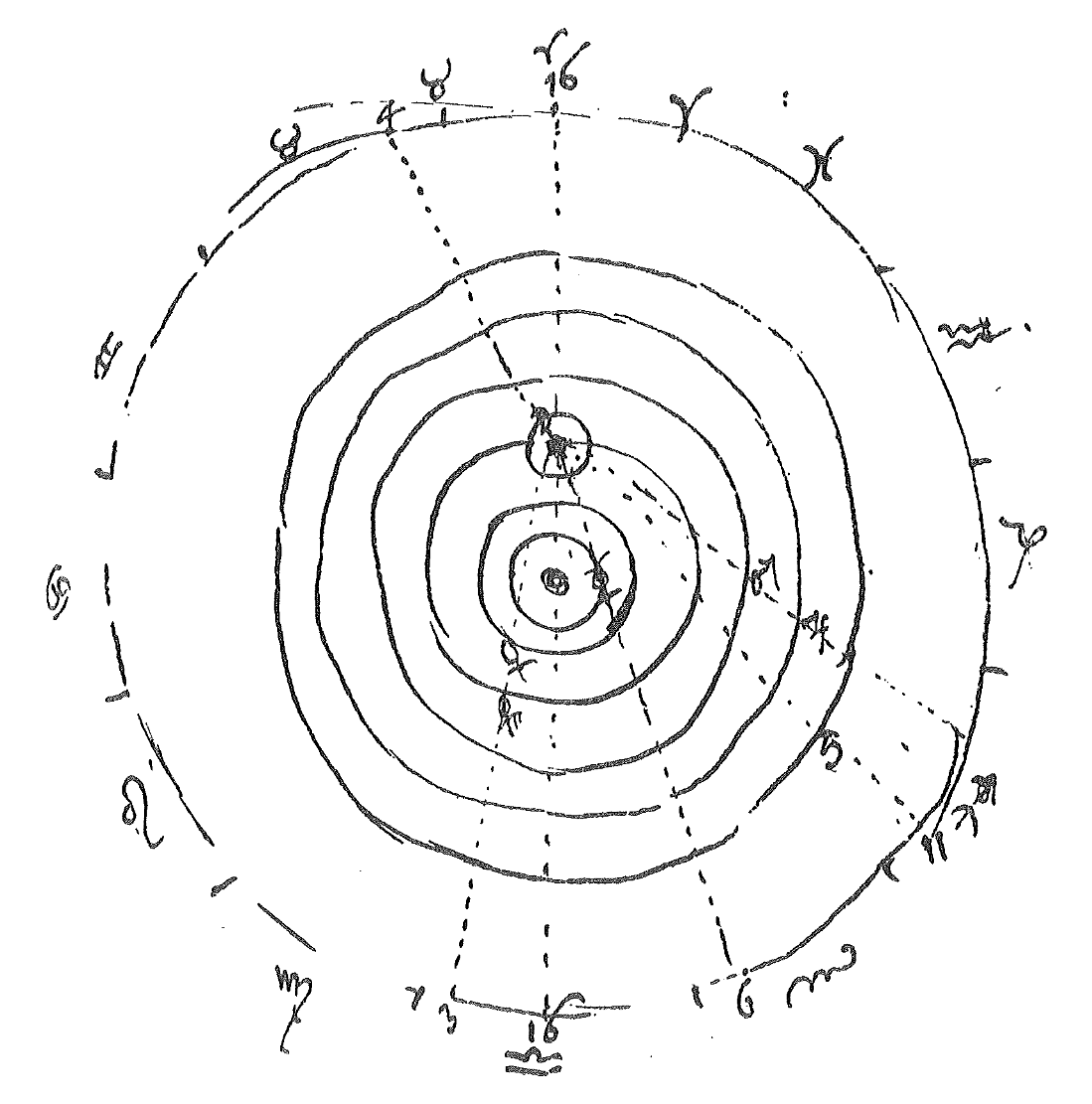}
\end{center}

\label{copernico}

\begin{center}
\includegraphics[width=0.6\linewidth]{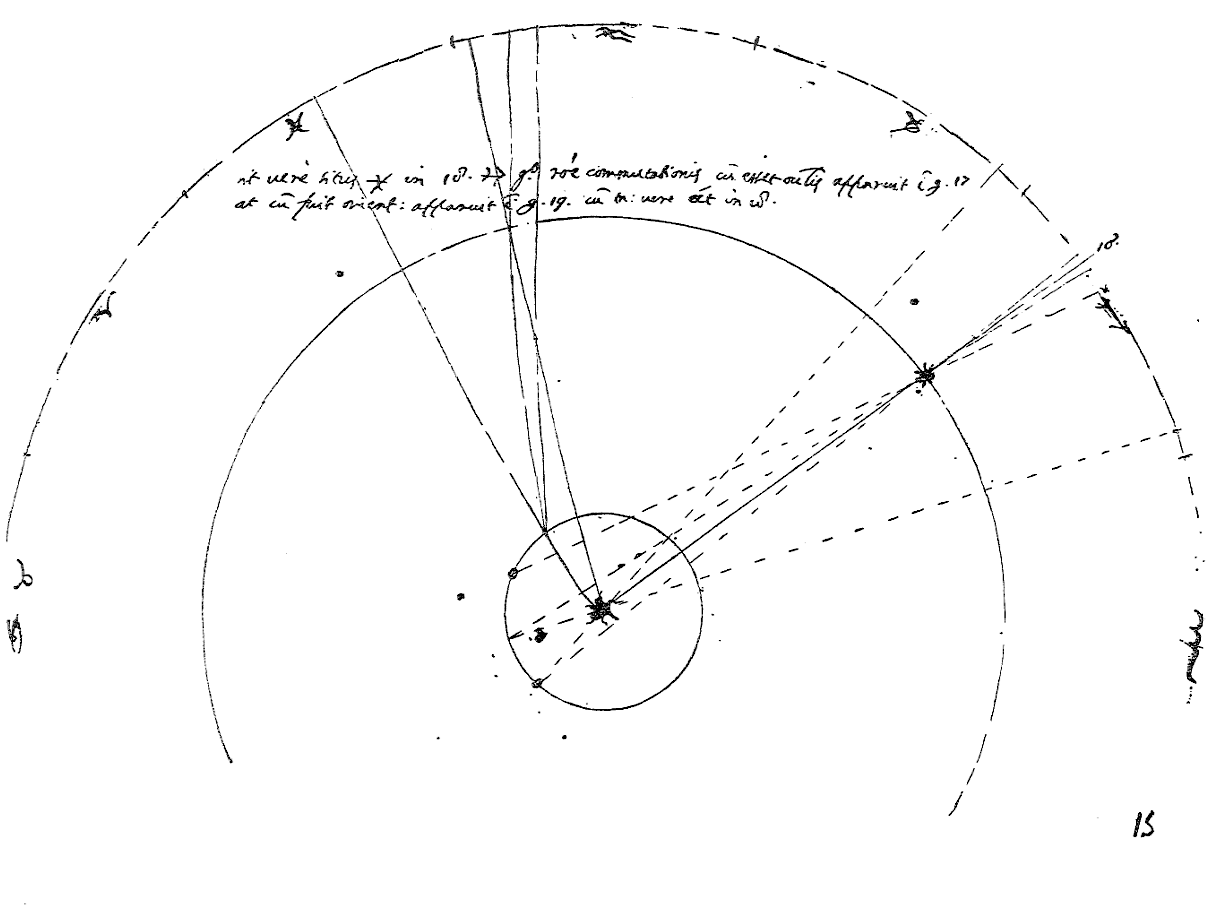}
\end{center}

Note the Copernican essence of the drawings, and the outer circle of the Zodiac with the signs. In the first drawing the Earth, third planet from the Sun which is the black point in the center, is indicated by a sign badly written, as if Galilei was afraid that his notes would fall into the wrong hands, but it is still between Mars and Venus and is the center of the observations -- a drawing
similar will be reproduced in the {\em Dialogue concerning the two chief world  systems} in 1632. In the second
you can see the seasonal variation of parallax. The drawings are not in the original first edition of the {\em Opere} \cite{opere}, but were added in an appendix to the second volume in later editions.

In addition to being more compelling than a treatise, as Plato had already shown and as prescribed in
numerous courses on rhetoric of Galilei’s times, a dialogue allows for circumventing formalism in certain
conclusions that Galilei was probably not able to develop rigorously: a dialogue allows its participants to
forego certain rigorous demonstrations and replace them with assumptions of sufficient plausibility, made
stronger by the use of common sense and of humour. In this
sense the technique of persuasion in this dialogue anticipates by three decades the two great Galilei's  dialogues: the {\em Dialogue concerning the two chief world  systems} (1632) and the {\em Discourses and mathematical demonstrations relating to two new sciences} (1638).


The present {\em Dialogue} appears as Galilei’s work in the second volume of the Italian national edition of
the {\em Opere} \cite{opere}. In this  edition it does not appear instead a poem in octaves (i.e., groups or ``stanzas'' of eight verses), by an unknown author,
published as an appendix to the first edition of 1605, and immediately replaced in the second edition. An English  translation of the {\em Dialogue}  appears in a book by Drake \cite{drake}; a commented translation in modern Italian is published in \cite{deasup}. Some (see \cite{daniele}, also for notes on the litterary production by Galilei) possibly attribute it to Antonio 
Querengo\footnote{The Paduan Antonio Querengo (or Querenghi, or Quarenghi, or Antuogno Squerengo in the dialectal mispronunciation) was a very learned diplomat and poet in Latin, a friend of Torquato Tasso. After the death in 1601 of Giovanni Vincenzo Pinelli, a good friend and sponsor of Galilei since his arrival in Padua, Querengo's house replaced Pinelli's as a mecca for all important visitors and a meeting place for local scholars and men of letters. Among these were Paolo Gualdo and Lorenzo Pignoria, who were also good friends of Galilei. Querengo continued to follow with interest the career of Galilei after he left Padua, and it is from his pen that we have the most amusing letters concerning the unfortunate campaign of Galilei in Rome in 1615-1616 to support the Copernican point of view.}, to whom the {\em Dialogue} is dedicated. 

The following  elements strengthen the attribution of the poem to Galilei:
\begin{itemize}
\item  Galilei had likely a strong participation to the {\em Dialogo de Cecco di Ronchitti}; being a poet (the full volume IX of the national edition of his {\em Opere}  \cite{opere}  is dedicated to his litterary activity), it appears strange that the poem in the appendix of the  {\em Dialogo} itself was written by somebody else, or at least that Galilei did not heavily rework it;
\item the style, including the verbal violence, is coherent with Galilei, and  there is  quite a strong attack to the Aristotelians;
\item as Milani observes in her 1992 translation of the {\em Dialogo} \cite{milani} (the translation of the text is done from the second edition \cite{bruzene2}), some rhymes are common with those of Galilei, and one stanza begins with the imperative ``Dunque", as in Galilei's reworking of Andrea Salvadori's {\em Canzone per le Stelle Medicee} (vol. IX of Galilei's {\em Opere}  \cite{opere});
\item a description of the method of parallax is appropriate for an astronomer, and a Copernican statement is present in the fourth {stanza} of the poem.
\end{itemize}
In any case, attribution always contains some degree of arbitrariness. 

Here follows the poem published in the appendix of the {\em Dialogue,} in the first edition of 1605.


\vskip 1 cm

\begin{center}
STANZE D'INCERTO CONTRA ARISTOTELE\\
PER LA  STELLA NUOVAMENTE APPARSA
\end{center}

\vskip 6mm

\begin{verse}
Che pi\`u vaneggi, o Stagirita stolto:\\ 
e puro il Cielo e ingenerabil credi?\\ 
Stella nuova, in lui fissa, il chiaro volto\\ 
discopre scintillando, e non la vedi?\\ 
O pi\`u che mai ne' primi errori involto\\ 
il senso neghi ed altre prove chiedi?\\ 
Il senso neghi, onde i principii certi \\
dicesti avere de le scienze aperti?\\



\vskip 4mm

Ma che nel Cielo sia la nata stella\\ 
stimerai forse non verace assunto;\\ 
e chiaro \`e pur che a questa gente e a quella\\si mostra fissa in un medesmo punto.\\ 
Vede ciascun che da la punta bella\\
di Sagittario ha il suo splendor disgiunto\\e quasi fugga di Scorpion la coda\\ 
s'erge tremante e sovra lei s'inchioda.\\ 

\vskip 4mm

Or s'ella \`e sotto il ciel, sotto la Luna,\\
s\`i lungi a quelle del sublime tetto,\\ 
come di un altro clima in parte alcuna\\ 
non si nasconde o cangia almen l'aspetto,\\ 
e al variar de' siti or sembra in una\\ 
or in altra figura aver ricetto,\\ 
anzi pur, come a noi torna mostrarse\\ 
nel mattutino ove la sera apparse?\\ 



\vskip 4mm

N\'e cessan qui le vere mie parole,\\ 
che a ragion ferme e nove ecco ritorno.\\
Dimmi, se volge la terrestre mole\\
ventidue mila miglia intorno intorno\\ 
e questa nova luce a par del Sole\\ 
tutta la gira in una notte e un giorno,\\
come tre ore e pi\`u sull'orizzonte\\ 
nostro nel tramontar mostr\`o la fronte?\\ 

\vskip 4mm

Esser non pu\`o che, s'ella umile \`e tanto, \\
cos\`i il passo allentar da noi sia vista, \\
ch\'e in breve spazio di mirarla il vanto \\
la retta cederebbe umana vista; \\
n\'e patria sempre assimigliar (se, quanto \\
da noi s'allunga, mole non acquista \\
e quella perde poi mentre s'appressa) \\
a l'occhio nostro la grandezza stessa. \\

\vskip 4mm

Ma poi ch'ella \`e fissa nel ciel sublime, \\
provan l'altezza sua s\`i ferme prove, \\
che come posta infra le stelle prime\\
a ragion seco si mostra e move; \\
n\'e d'esser pu\`o che in altre parti o in ime\\
varia di mole o sito, ella si trove, \\
ch\'e nulla face il mutar loco in terra \\
all'ampiezza del Cielo, in cui si serra. \\

\vskip 4mm

Dunque di cecit\`a squarciando il panno\\
omai la lingua sciogli in vere note \\
e se credesti con tuo scorno e danno \\
dianzi immutabil le soperne rote,\\ 
or che fiamma novella apre l'inganno\\
conosci il Ciel che generar si puote,\\ 
e grazie rendi a la Natura madre, \\
ond'hai luci del ver tanto leggiadre.\\ 


\end{verse}
\vskip 0.8  cm

This is my translation:

\vskip 1  cm

\begin{center}
STANZAS BY AN UNKNOWN AUTHOR TO THE NEWLY APPEARED STAR\\
AGAINST ARISTOTLE
\end{center}

\begin{verse}

What raving, o foolish Stagirite,\\
your belief that creations are forbidden in Heaven!\\
A newborn star, in it fixed, its clear face\\
 shows glittering, and you don't see it?\\
Or  you deny sense more than ever\\
getting stuck in your first mistakes, and ask for other proofs?\\
But how to deny this simple evidence\\
when all your science was based on sense?\\

\vskip 4mm

That in Heaven the star was  born \\
maybe you think  not to be true;\\
though every nation sees it  \\
fixed in the same point up on the sky.\\
Everybody sees it in the beautiful location\\
where Sagittarius has its splendor separated\\
and as if from  Scorpion's tail  escaping\\
rises trembling, and  it is nailed over it.\\

\vskip 4mm

Now if it be under the Heavens, under the Moon,\\
so far from the stars of the sublime roof,\\
as of another climate, how it does not partly hide,\\
 or at least changes its aspect, \\
and in a variety of sites seems now \\
in the same constellation to have shelter?\\
Indeed, how does it return to us in the morning\\
 where it appeared in the evening?\\

\vskip 4mm
 
Nor here cease my true words,\\
which  firm and new now return. \\
Tell me, if the Earth  \\
turns twenty-two thousand miles around\\
and this new light, like for the Sun\\
turns it all round in a night and a day,\\
 how could it for three hours and more\\
  on our horizon  show its face at sunset?\\

\vskip 4mm

It cannot be that, if it is so humble,\\
it sets its pace as to be seen, \\
 since in a short  time\\
 it would become invisible;\\
nor it could look always similar, unless, as it lengthens from us,\\
 it does  acquire mass\\
and then loses it while it draws near\\
to look the same in bulk at every place.\\

\vskip 4mm

But since it is fixed in the sublime sky,\\
 Its height proves so firmly, that \\
as placed between the first stars\\
naturally  shows itself there and moves accordingly;\\
nor  it could be found in lower  parts\\
   varying in size or site,\\
since  its place on Earth does not change\\
given the enormous circumference of the Heavens.\\

 \vskip 4mm

Therefore,  ripping the cloth blinding you,\\ 
now let your tongue melt into true notes:\\
if you believed, at your scorn and harm\\
 the upper wheels to be immutable,\\
now that this new flame shows  your mistake,\\
you know that Heaven  can be generated,\\
 and  thank  Mother Nature, \\
who sent you such graceful light of truth.\\

\end{verse}
\vskip 4mm
 A second edition of the  {\em Dialogo de Cecco di Ronchitti} was published later the same year in Verona \cite{bruzene2}. The poem was largely amended, with some uncertainties in the poetic language. In addition, the dangerous Copernican allusion to Earth rotation in the fourth stanza was removed, and some criticisms to Aristotle were softened and became
 ``politically correct'' (for example, in the first verse the sentence ``stolto Stagirita'', where ``stolto'' literally means ``fool'', ``foolish'', or even ``stupid'' \cite{cambridge}, became ``per altro Stagirista saggio'', i.e., ``otherwise sage Stagirite''). This new version, which I leave here untranslated, when compared to other poems by Galilei,  lacks the usual verbal aggressivity.
 
\vskip 4mm
\begin{verse}

Dunque, o per altro Stagirita saggio, \\
vaneggi ancora, e l'alte rote credi \\
immutabili e pure? Un novo raggio \\
tra quei zaffiri \`e nato, e non lo vedi? \\
O pur facendo a te medesmo oltraggio \\
il senso neghi, ed altre prove chiedi? \\
Il senso neghi, onde i principii certi \\
dicesti aver de le scienze aperti? \\

\vskip 4mm

Ma che in Ciel sia la generata stella \\
stimerai forse tu pensier fallace; \\
e chiaro \`e pur che questa gente e quella \\
un solo osserva in lei sito verace; \\
vede tra' segni luminosi ch'ella \\
di Sagittario al dardo aggiunge face, \\
e quasi fugga di Scorpion la coda, \\
scintilla e s'alza e sovra lei s'inchioda. \\

\vskip 4mm

Or quando, al tuo parer, sotto la Luna \\
resti l'obliquo suo camin ristretto, \\
come non mai s'asconde, e non s'imbruna \\
a' varii Climi opposta, o cangia aspetto? \\
E non rassembra a noi dal Cielo in una, \\
ad altri in altra parte haver ricetto? \\
Anzi come il mattin torna mostrarse \\
fissa nel punto ove la sera apparse? \\

\vskip 4mm

Aggiungo: e come all'or, che parte o riede \\
e si mostra or d'appresso, et or distante \\
al Zenit nostro, non mai muta o cede \\
la figura, e la luce ogn'or tremante? \\
Come in lungo passaggio non la vede, \\
e meno o pi\`u distinta, occhio costante? \\
E non cresce e non scema, al guardo esposta, \\
qual'or, girando, s'avvicina e scosta? \\

\vskip 4mm

Meglio dir\`o: s'ella il suo corso stende \\
pel vicino sentier de gli Elementi, \\
come sopra di noi tant'ore splende \\
quante del sommo Giro i lumi ardenti? \\
Ed essi regolato il moto apprende, \\
né vien che pi\`u s'affretti, o pi\`u s'allenti,\\ 
come l'infimo raggio col supremo \\
ne va del par da l'uno a l'altro estremo?\\

 \vskip 4mm
 
Esser non pu\`o che in cerchio angusto e basso \\
 de' corpi eccelsi al gran viaggio assista, \\
e mentre tien su l'Orizonte il passo \\
di mille aspetti al variar resista, \\
ché d'umil Cerchio una gran parte il crasso \\
globo ben toglie a la terrena vista; \\
ma la spera stellata ei non ingombra, \\
in cui riguardo egli d'un punto \`e l'ombra. \\

\vskip 4mm

Or questi effetti indubitati e veri \\
son del nostro saper fondate prove, \\
poiché la stella unita a i segni alteri \\
tra quei scintilla, e si dimostra, e move; \\
né da gl'Indi scorrendo a i lidi Iberi, \\
fia chi di mole o forma altra la trove, \\
ché simil variet\`a non cape il senso, \\
né la distanza di quell'Orbe immenso.\\ 

\vskip 4mm

Dunque, o Maestro di color che sanno, \\
sgombra le nubi omai da l'intelletto, \\
e se credesti, con tuo scorno e danno,\\ 
dianzi immutabil quel soperno oggetto, \\
or che fiamma novella apre l'inganno \\
fatto il confessa, e a corruttion soggetto, \\
e rendi gratie a la Natura Madre, \\
ond'hai luci del ver tanto leggiadre. \\

\end{verse}

\subsection*{Acknowlegedments}
I thank Selenia Broccio and  Marianna Giannicolo for their help, and Sandro Bettini, Michele Camerota, Luisa Cifarelli, Ivano Paccagnella and Bill Shea for their comments and suggestions (Bill also for many pleasant discussions). 

This article is dedicated to the 800th anniversary of the University of Padova, and has been written in the framework of the celebrations. {\em Universa universis patavina libertas.}

\end{document}